\documentclass[sigconf,screen,nonacm]{acmart}
\AtBeginDocument{%
  }

\setcopyright{acmlicensed}
\copyrightyear{2026}
\acmYear{2026}
\acmDOI{XXXXXXX.XXXXXXX}
\acmConference[FutureHCI '26]{The Future of HCI Workshop}{August 17--18, 2026}{Blacksburg, VA, USA}
\acmISBN{978-1-4503-XXXX-X/2026/06}

\usepackage{rotating}
\usepackage{array}
\usepackage{booktabs}
\usepackage[table]{xcolor}
\usepackage{tabularx}
\usepackage{hyperref}
\usepackage{makecell}

\begin{document}

\title{The Capturing and Logging Ecological Virtual Experiences and Reality (CLEVER) — Job Simulator Dataset}


\author{Qidi J. Wang}
\orcid{0009-0005-7372-5891}
\affiliation{%
  \institution{Virginia Tech}
  \city{Blacksburg}
  \state{Virginia}
  \country{USA}
}
\email{qidiwang@vt.edu}

\author{Xiaozheng Wang}
\orcid{0009-0004-6886-294X}
\affiliation{%
  \institution{Virginia Tech}
  \city{Blacksburg}
  \state{Virginia}
  \country{USA}
}
\email{xzwang@vt.edu}

\author{Akhilesh M. Anand}
\orcid{0009-0007-6327-8762}
\affiliation{%
  \institution{Virginia Tech}
  \city{Blacksburg}
  \state{Virginia}
  \country{USA}
}
\email{anandakhilesh29@vt.edu}

\author{Veera V. Pala}
\orcid{0009-0005-7152-2118}
\affiliation{%
  \institution{Virginia Tech}
  \city{Blacksburg}
  \state{Virginia}
  \country{USA}
}
\email{veera@vt.edu}

\author{Rohan V. Penmetsa}
\orcid{0009-0000-5131-573X}
\affiliation{%
  \institution{Virginia Tech}
  \city{Blacksburg}
  \state{Virginia}
  \country{USA}
}
\email{rohanpenmetsa@vt.edu}

\author{Ryan P. McMahan}
\orcid{0000-0001-9357-9696}
\affiliation{%
  \institution{Virginia Tech}
  \city{Blacksburg}
  \state{Virginia}
  \country{USA}
}
\email{rpm@vt.edu}

\renewcommand{\shortauthors}{Wang et al.}

\begin{abstract}
  Virtual reality (VR) motion tracking and interaction data has become increasingly recognized as valuable for machine learning experiments for a variety of purposes, including predicting user identities, predicting user attributes like gender and age, predicting retention and learning, and more. However, there exist a limited number of publicly accessible VR motion datasets. In this paper, we present a new open-source dataset of 95 participants playing the SteamVR game \textit{Job Simulator}. Additionally, we review existing datasets, detail our study procedure, describe our data collection process, list attributes of our dataset, and suggest future work, impact, and applications.
\end{abstract}

\ccsdesc[500]{Human-centered computing~Virtual reality}
\ccsdesc[300]{Human-centered computing~Empirical studies in HCI}
\ccsdesc[100]{Human-centered computing~User studies}

\keywords{Virtual reality, open dataset, machine learning}
\begin{teaserfigure}
  \includegraphics[width=\textwidth]{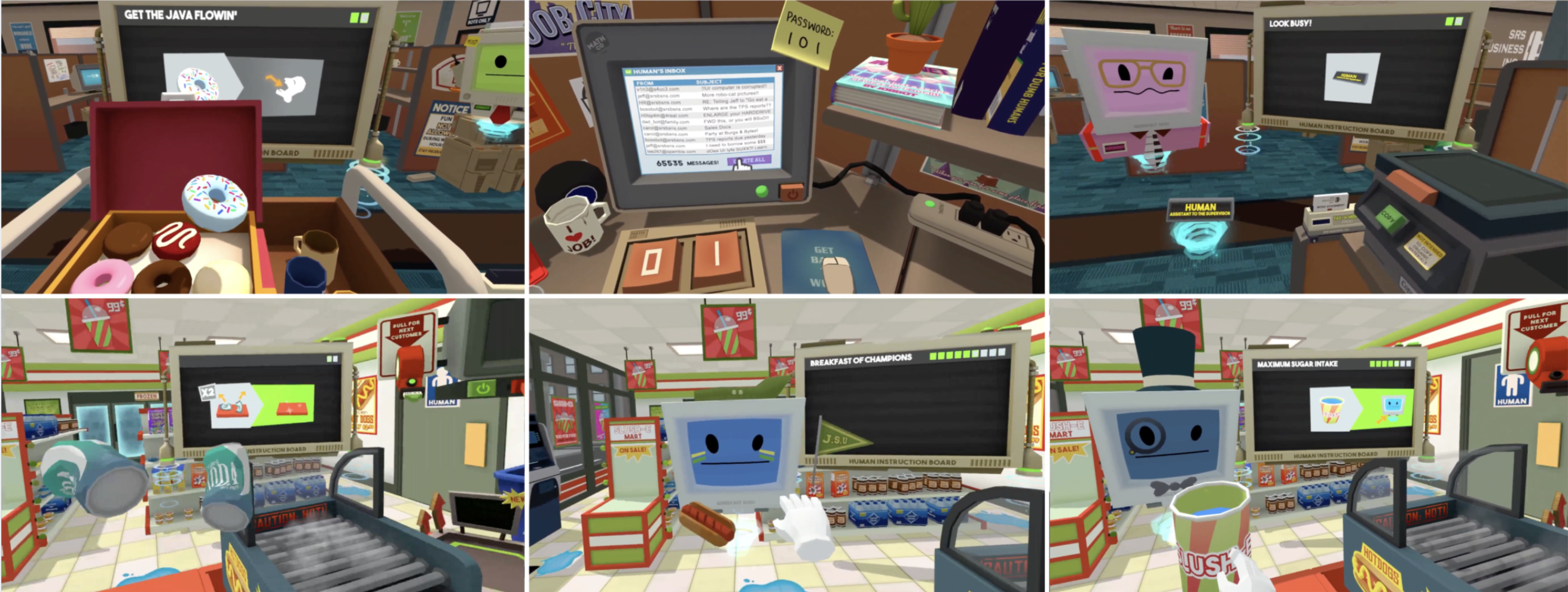}
  \caption{\textit{Job Simulator} tasks in the \textit{Office Worker} and \textit{Store Clerk} domains. From top left corner to bottom right corner: ``Get the Java Flowin'" (\textit{Office Worker} Task 1), ``Wake Up Computer" (\textit{Office Worker} Task 2), ``Look Busy!" (\textit{Office Worker} Task 3), ``Open for Business" (\textit{Store Clerk} Task 1), ``Breakfast of Champions" (\textit{Store Clerk} Task 2), ``Maximum Sugar Intake" (\textit{Store Clerk} Task 3).}
  \Description{\textit{Job Simulator} Tasks}
  \label{fig:teaser}
\end{teaserfigure}


\maketitle

\begin{table*}[!t]
\caption{%
Comparison of related work and our work in terms of the number of users, number of devices per user, number of applications, number of domains per user, number of unique tasks, number of sessions per user, applications, and whether the dataset is publicly accessible. We use ? to symbolize values that varied or were not reported in the original work. For the Public Accessibility column, a $\varnothing$ denotes a dataset that is not stated to be openly available, $\dagger$~indicates an open dataset that is referenced but for which no public link was provided, and $\times$ indicates a dataset with a public link but is inaccessible. Datasets with fewer than 50 users are indicated with a gray background.
}
\label{tab:relworks_dataset}
  \normalsize%
  \centering%
\setlength{\tabcolsep}{4pt}
\renewcommand{\arraystretch}{1.3}
\renewcommand{\theadfont}{\bfseries}

\begin{tabularx}{\linewidth}{p{2.8cm}|r r r r r r | X |c}
\textbf{Reference}
  & {\rotatebox{90}{\textbf{\# Users}}}
  & {\rotatebox{90}{\textbf{\# Devices/User}}}
  & {\rotatebox{90}{\textbf{\# Applications}}}
  & {\rotatebox{90}{\textbf{\# Domains/User}}}
  & {\rotatebox{90}{\textbf{\# Unique Tasks}}}
  & {\rotatebox{90}{\textbf{\# Sessions/User}}}
  & {\textbf{Applications}}
  & {\rotatebox{90}{\textbf{Public Accessibility}}}
\\ \hline

Nair et al. \cite{Nair2023b}          
  & 105{,}852 & 1 & 2 & 1 & ?
  & ?
  & \textit{Beat Saber} or \textit{Tilt Brush}
  & \checkmark \\

Moore et al. \cite{Moore2025}          
  & 108 & 1 & 1 & 1 & 2
  & 2
  & Assembly
  & \checkmark \\

Rack et al. \cite{Rack2023} 
  & 71 & 1 & 1 & 1 & 2
  & 2
  & \textit{Half-Life: Alyx}
  & \checkmark \\

Baldoni et al. \cite{baldoni_questset_2024}       
  & 60 & 1 & 4 & 2 & 4--5
  & 2
  & \textit{Beat Saber} and \textit{Cooking Simulator} or
  & \checkmark \\

  &  &  &  &  &  &
  & \textit{Medal of Honor: Above and Beyond} and \textit{Forklift Simulator}
  &  \\

\rowcolor{gray!15}
Schach et al. \cite{Schach2026}        
  & 49 & 1 & 5 & 5 & ?
  & 5
  & \textit{Half-Life: Alyx}, \textit{Superhot VR}, \textit{Beat Saber}, \textit{Synth Riders}, Social VR
  & \checkmark \\
  
\rowcolor{gray!15}
Rack et al. \cite{Rack2024c}         
  & 48 & 1 & 1 & 1 & 4
  & 2
  & Motion Password
  & \checkmark \\

\rowcolor{gray!15}
Moore et al. \cite{Moore2023}          
  & 45 & 1 & 1 & 1 & 2
  & 2
  & Assembly Tasks 
  & \checkmark \\

\rowcolor{gray!15}
Miller et al. \cite{miller_within-system_2020} 
  & 41 & 3 & 1 & 1 & 1
  & 6
  & Throwing a Ball
  & $\varnothing$ \\

\rowcolor{gray!15}
Thiel \& Steed \cite{ThielSteed2022}    
  & 20 & 1 & 7 & 1--4 & ?
  & 1--4
  & \textit{Beat Saber}, \textit{Clash of Chefs}, \textit{Half-Life: Alyx}, \textit{Job Simulator}, \textit{Pistol Whip}, \textit{Space Pirate Trainer}, \textit{The Lab}
  & $\dagger$ \\

\rowcolor{gray!15}
Liebers et al. \cite{Liebers2024} 
  & 16 & 2 & 1 & 1 & 8
  & 2
  & Multi-Interface Interaction
  & \checkmark \\

\rowcolor{gray!15}
Liebers et al. \cite{Liebers2021}        
  & 16 & 1 & 2 & 2 & 2
  & 4
  & Bowling and Archery
  & \checkmark \\

\rowcolor{gray!15}
Wen et al. \cite{Wen2024}            
  & 16 & 1 & 100 & ? & ?
  & ?
  & 100 games spanning 10 genres (e.g. \textit{Beat Saber}, \textit{Cartoon Network}, etc.)
  & $\times$ \\

\rowcolor{gray!15}
Liebers et al. \cite{Liebers2023}        
  & 15 & 1 & 1 & 1 & 1
  & 4--16
  & \textit{Beat Saber}
  & \checkmark \\ \hline

\textbf{Ours}
  & \textbf{95} & \textbf{2} & \textbf{1} & \textbf{2} & \textbf{6}
  & \textbf{5}
  & \textbf{\textit{Job Simulator}}
  & \textbf{\checkmark} \\

\end{tabularx}
\end{table*}

\section{Introduction}

There has been a growing body of work exploring what can be inferred from VR motion and interaction data. Researchers have explored the feasibility of user identification and authentication \cite{id1, miller2023, id2, id3, id4}, gender identification \cite{Wang2024, nair_exploring_2023}, as well as predicting simulator sickness \cite{cybersickness}, learning \cite{extract, explore}, cognitive load \cite{workload},  and more. 

As more machine learning experiments are conducted, there arises a need for more publicly available VR motion tracking datasets. There exist open datasets that involve throwing a ball \cite{miller_within-system_2020}, bowling and shooting arrows \cite{Liebers2021}, playing \textit{Beat Saber} \cite{Nair2023b, baldoni_questset_2024, Liebers2023, ThielSteed2022, Wen2024}, playing \textit{Half-Life: Alyx} \cite{Rack2023}, assembling pipe structures \cite{Moore2023, Moore2025}, and more (See Table \ref{tab:relworks_dataset}). Historically, researchers had to develop their own applications or collection techniques, which could be tedious and time-consuming. However, Segarra Martinez et al. have recently presented CLOVR (Capturing and Logging OpenVR data), an open-source recording toolkit that can be overlaid on top of existing, closed-source SteamVR games and applications, which serves to assist researchers in the data collection process \cite{Martinez2024}. 

In this paper, we present the open-source Capturing and Logging Ecological Virtual Experiences and Reality (CLEVER) — Job Simulator dataset \footnote{\url{https://huggingface.co/datasets/xraijobsimulator/CLEVER_Job_Simulator}}, which was captured using CLOVR. Our dataset consists of 95 participants (41 females, 53 males, 1 other) and includes their motion and interactions, Fidelity-based Presence Scale (FPS) and Simulator Sickness Questionnaire (SSQ) responses, videos from the HMD's perspective, and project metadata. In this paper, we survey prior datasets, outline our study methodology, describe how data was collected, detail attributes of our dataset, and suggest future work and impacts.

\section{Related Work}

We have compiled a list of existing VR motion tracking datasets in Table \ref{tab:relworks_dataset}. These are compared according to the number of users, number of devices per user, number of applications, number of domains per user, number of unique tasks, number of sessions per user, applications, and whether the dataset is publicly accessible. ``Users" are the individuals who underwent the experimental procedure and whose data was recorded. ``Devices" refers to distinct VR systems, such as the Meta Quest 3S and VIVE XR Elite, that may or may not be manufactured by different companies. An ``application" is a program or game that provides the setting for the tasks the user can perform, like \textit{Half-Life: Alyx} or \textit{Beat Saber}. ``Domains" describe one or more categories within the application that tasks belong to. For example, Moore et al.'s \textit{Full-scale Assembly Simulation Testbed (FAST)} datasets belong in the assembly domain \cite{Moore2023, Moore2025}, Rack et al.'s \textit{Who is Alyx?} dataset belongs in the first-person shooter domain \cite{Rack2023}, and datasets with \textit{Beat Saber} belong in the rhythm game domain \cite{Nair2023b, baldoni_questset_2024, Schach2026, ThielSteed2022, Wen2024, Liebers2023}. ``Tasks" are the unique activities users complete with well-defined beginning and end points within a domain. Finally, we define ``sessions" to begin when the user puts on the headset and end when the user takes it off. Our definition sometimes differs from the definition of other authors. For example, Schach et al.'s dataset involves a single procedure that consists of five applications \cite{Schach2026}. We reported this dataset as having five sessions because the user can take off their headset between applications.   

\subsection{Users}
These datasets range from small to large. Nair et al.'s dataset BOXRR-23 is very large, consisting of 105,852 users. Data was collected from real \textit{Beat Saber} and \textit{Tilt Brush} players \cite{Nair2023b}. It is worth noting that these users typically played \textit{Beat Saber} rather than both. While the magnitude of this dataset is impressive, it is not easily accessible due to the amount of storage it requires to be downloaded onto a local machine and the unique data format it uses, Extended Reality Open Recording (XROR). The datasets of Moore et al. \cite{Moore2025}, Rack et al. \cite{Rack2023}, and Baldoni et al. \cite{baldoni_questset_2024} are large, with 108, 71, and 60 users respectively. Many range in the 40s \cite{Schach2026, Rack2024c, Moore2023, miller_within-system_2020}. The rest are small, with 20 or fewer users \cite{ThielSteed2022, Liebers2021, Liebers2023, Liebers2024, Wen2024}. We applied a gray background to the datasets with fewer than 50 users because scikit-learn recommends that at least 50 samples are obtained before applying machine learning algorithms \cite{sklearn}. Our dataset consists of 95 individuals, making it the third-largest dataset listed here.

\subsection{Devices Per User}
The majority of datasets recorded motion via a single device \cite{Nair2023b, Moore2023, Moore2025, Rack2023, baldoni_questset_2024, Schach2026, Rack2024c, ThielSteed2022, Liebers2021, Liebers2023, Wen2024}. However, there are a few exceptions: Miller et al. used three devices (Oculus Quest, HTC Vive, and HTC Vive Cosmos) \cite{miller_within-system_2020} and Liebers et al. used two devices (Microsoft HoloLens 2, Meta Quest 2) \cite{Liebers2024}. We also provided one of the few datasets that used more than one device by including the Meta Quest 3S and the VIVE XR Elite.  

\subsection{Applications}
Most of these datasets have one application, but some have more. Nair et al. pulled data from \textit{Beat Saber} and \textit{Tilt Brush} \cite{Nair2023b}. Baldoni et al. utilized four applications: \textit{Beat Saber}, \textit{Cooking Simulator}, \textit{Forklift Simulator}, and \textit{Medal of Honor: Above and Beyond} \cite{baldoni_questset_2024}. Schach et al. had each of their participants experience all five applications in their study procedure: \textit{Half-Life: Alyx}, \textit{Superhot VR}, \textit{Beat Saber}, \textit{Synth Riders}, and Social VR \cite{Schach2026}. Thiel \& Steed's participants could choose from seven applications: \textit{Beat Saber}, \textit{Clash of Chefs}, \textit{Half-Life: Alyx}, \textit{Job Simulator}, \textit{Pistol Whip}, \textit{Space Pirate Trainer}, and \textit{The Lab} \cite{ThielSteed2022}. Liebers et al. used bowling and archery as their two applications \cite{Liebers2021}. Wen et al. selected 100 games across 10 genres for their participants to experience \cite{Wen2024}. Our dataset used one application: \textit{Job Simulator}. 

\subsection{Domains Per User}
Users typically experienced one domain, with some exceptions. Baldoni et al. split their participants into two groups \cite{baldoni_questset_2024}. One group played \textit{Beat Saber} and \textit{Cooking Simulator} while the other group played \textit{Forklift Simulator} and \textit{Medal of Honor: Above and Beyond}. Regardless of group, each participant experienced two domains by playing two games. Each of Schach's 5 applications represented a separate domain \cite{Schach2026}. Thiel \& Steed gave their participants freedom to choose which VR games they wanted to play, so the domains per user varied between 1 and 4 \cite{ThielSteed2022}. Bowling and archery are two distinct domains in Liebers et al.'s work \cite{Liebers2024}. The number of domains each user experienced varied in Wen et al.'s dataset because they did not specify how many applications each user experienced \cite{Wen2024}. Our dataset consisted of two domains within \textit{Job Simulator}: \textit{Office Worker} and \textit{Store Clerk}.

\subsection{Unique Tasks}
The number of unique tasks varied by dataset, but most had more than one. The users in Baldoni et al.'s dataset experienced either four or five tasks, depending on which group they were in \cite{baldoni_questset_2024}. Thus, their dataset had nine unique tasks in total. Participants in Rack et al.'s \textit{Motion Passwords} dataset each traced four words, making four tasks, but the total number of words in this dataset is unknown as two of the four words were random \cite{Rack2024c}. Liebers et al. had participants interact with eight user interfaces elements, making for eight tasks \cite{Liebers2024}. Our dataset consisted of three unique tasks for \textit{Office Worker} and three for \textit{Store Clerk}, making six unique tasks total, as shown in Figure \ref{fig:teaser}.

\subsection{Sessions Per User}
Most of the users represented in these datasets experienced more than one session. That is, they put on and took off the headset multiple times throughout the course of the experimental procedure. Our dataset consists of five sessions per user. Three sessions are completed in the first device (either the Meta Quest 3S or the VIVE XR Elite), and two are completed in the other device.

\subsection{Public Accessibility}
Almost every dataset presented here is open-source and publicly accessible, except for Miller et al.'s dataset of ball-throwing \cite{miller_within-system_2020}. Thiel \& Steed's dataset is referenced as publicly accessible in their paper, but no public link was provided \cite{ThielSteed2022}. Wen et al. provide a link to their dataset, VR.net, but the link cannot be opened \cite{Wen2024}. Our dataset is open-source and available on HuggingFace.

\begin{figure*}
  \includegraphics[width=\textwidth]{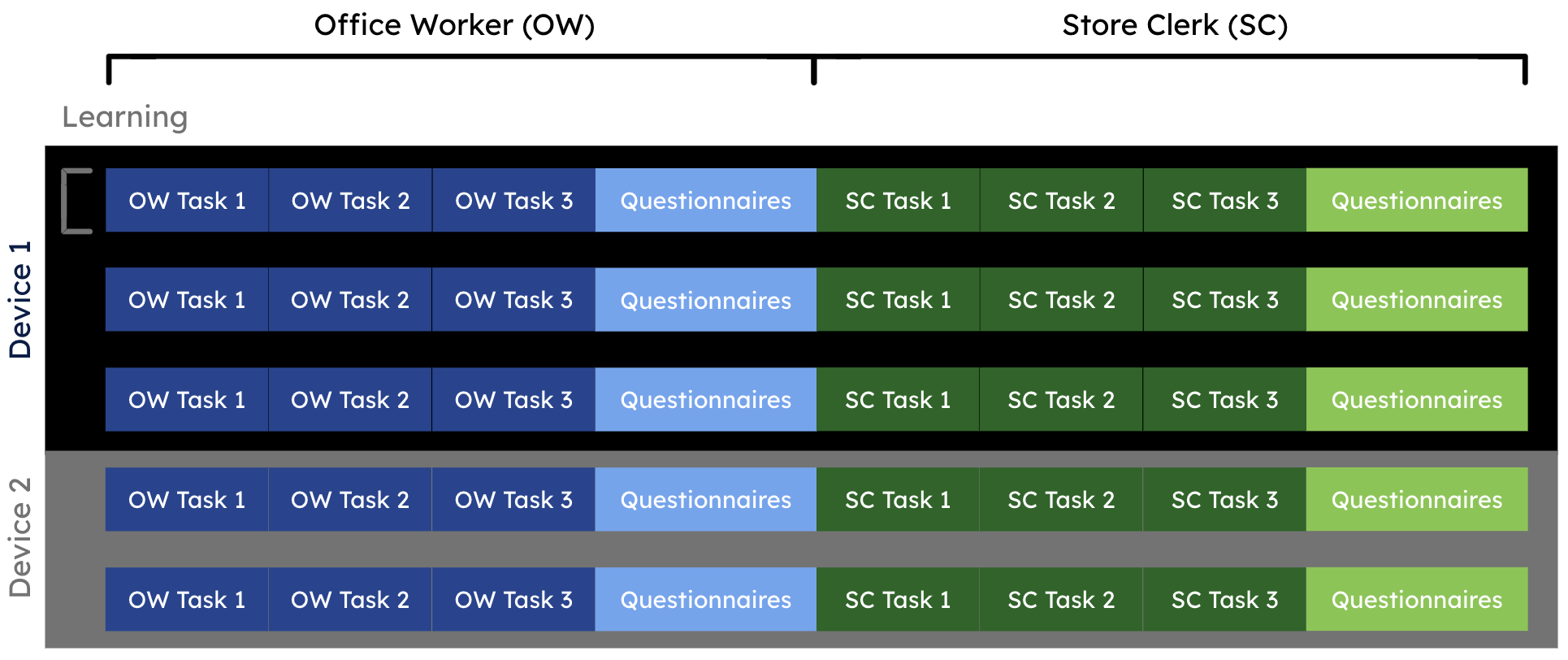}
  \caption{Study Procedure Design.}
  \label{fig:studystructure}
\end{figure*}

\section{VR Game: \textit{Job Simulator}}

\textit{Job Simulator} consists of four occupations for users to play through, each with different settings and tasks: \textit{Office Worker}, \textit{Store Clerk}, \textit{Auto Mechanic}, and \textit{Gourmet Chef}. Specifically, we focused on the \textit{Office Worker} and \textit{Store Clerk} domains and included the first three tasks from each domain, resulting in six tasks per session that capture a variety of VR interaction behaviors. 
\subsection{Office Worker}

The \textit{Office Worker} domain takes place in an office cubicle setting. The first task of \textit{Office Worker} is titled ``Get the Java Flowin'", and the player must fill up a coffee mug, drink the coffee, and eat a donut. The second task of \textit{Office Worker} is titled ``Wake Up Computer", and the player must plug in the computer tower and monitor, enter a login password, and delete emails. The third task, ``Look Busy!''. is comparatively open-ended, allowing participants to perform improvised interaction with components in the VR environment until the Supervisor Bot promotes the player and hands them a desk nameplate.

\subsection{Store Clerk}

The \textit{Store Clerk} domain takes place in a convenience store setting. The first task of \textit{Store Clerk} is ``Open for Business", and the player must clear the counter of two empty soda cans as well as turn on the security camera. The second task of \textit{Store Clerk} is titled ``Breakfast of Champions", and the player must scan the customer's chips and grill a hot dog. The third task of \textit{Store Clerk} is titled ``Maximum Sugar Intake", and the player must scan the customer's candy bars as well as make a jumbo-sized slushie.

\section{Data Collection Design}

\subsection{Materials}

We collected data using two VR systems: a Meta Quest 3S and a VIVE XR Elite. Each headset was connected to a dedicated Alienware 18 laptop to maintain a consistent experimental setup and enable efficient transitions between devices. 

\subsection{Participants}
A total of 95 participants were recruited through university email lists, word-of-mouth, and printed flyers. Each interested individual completed a pre-screening survey to evaluate whether they met all eligibility criteria before scheduling a 90--minute study session.

\subsection{Procedure}

The following procedure was approved by the Virginia Tech Institutional Review Board (IRB).

The procedure consisted of experiencing two occupations in \textit{Job Simulator}: \textit{Office Worker} and \textit{Store Clerk}. For each of five sessions, the participant would play through the first three tasks in \textit{Office Worker}, navigate back to the lobby to take two questionnaires, play through the first three tasks in \textit{Store Clerk}, and then navigate back to the lobby again to take the same two questionnaires (Figure \ref{fig:studystructure}). 

There were five individuals who ran participants for this dataset, who we term ``study runners". Each iteration of the procedure was handled by one of the study runners. 

The study runner would set up the devices before any participant entered the room, which involved configuring the virtual boundary for both the Meta Quest 3S and the VIVE XR Elite. For the Quest 3S, the runner drew a roomscale boundary along the edges of the wall of the study testing room. For the VIVE XR Elite, the runner set the boundary to be a stationary circle of radius 3m, which provided sufficient room for the user to move freely and complete their tasks without hitting walls in the real world. We chose to use a stationary boundary for the VIVE XR Elite instead of a roomscale boundary due to inconsistent boundary size estimates in SteamVR, which was an issue unique to the VIVE XR Elite. When we drew roomscale boundaries in the VIVE XR Elite, the SteamVR play area dimensions would sometimes be greater than 2.5m × 1.875m and sometimes be less. This dimensional threshold is crucial because play areas smaller than it would trigger the stationary version of \textit{Job Simulator} to appear in the VIVE while the the roomscale version of \textit{Job Simulator} consistently appeared in the Quest. Such inconsistency in game versions was unacceptable for our study, so we chose to keep the VIVE XR Elite boundary a stationary circle of radius 3m, which consistently resulted in a 6m x 6m play area that ensured the roomscale version of \textit{Job Simulator} appeared in both devices. Additionally, we understood that floor height could affect features downstream, including head position y values, which encode estimates for user height. Thus, the study runner would reset the floor height in both devices and check that the virtual controllers touched the virtual floor when the real controllers were set on the real ground. We also used a pulley system attached to the ceiling to suspend the cable connecting the headset to the corresponding laptop to prevent tripping. 

The study runner configured CLOVR in preparation for the session. We used CLOVR to capture motion data, questionnaire responses, game audio, microphone audio, video of gameplay from the user’s perspective as shown in SteamVR’s VR View window set to ``Both Eyes – Right Eye Dominant”, and project metadata. We also input the participant’s randomly assigned PID number and session number (1-5) into the CLOVR graphical user interface.

When the participant came in for their session, the study runner greeted them and asked them to review the printed consent document if they had not already. The study runner then verbally reviewed key points from the consent document, including the purpose of the study, procedures, notice of data collection and anonymization, and permission to withdraw at any time. Once the participant confirmed they understood and agreed and did not have any questions, both the participant and the study runner signed the consent form. 

Next, the study runner directed the participant to complete a background survey, which collected demographic information such as gender, age, hand dominance, prior video game experience, and prior virtual reality experience.

\begin{figure}
  \includegraphics[width=0.35\textwidth]{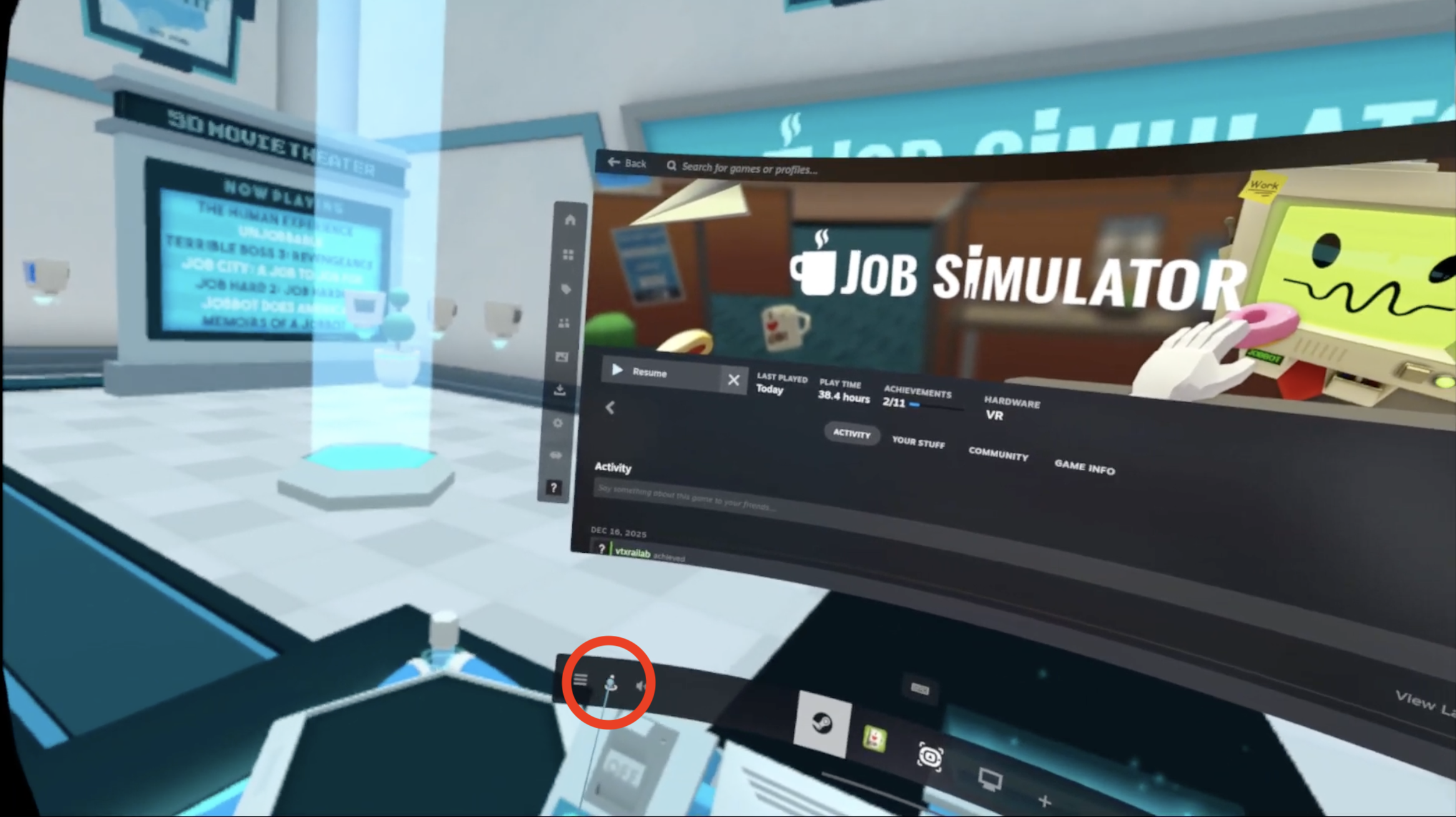}
  \caption{Recentering Button on Bottom Left of Menu Panel.}
  \label{fig:recentering}
\end{figure}

\begin{figure}
  \includegraphics[width=0.35\textwidth]{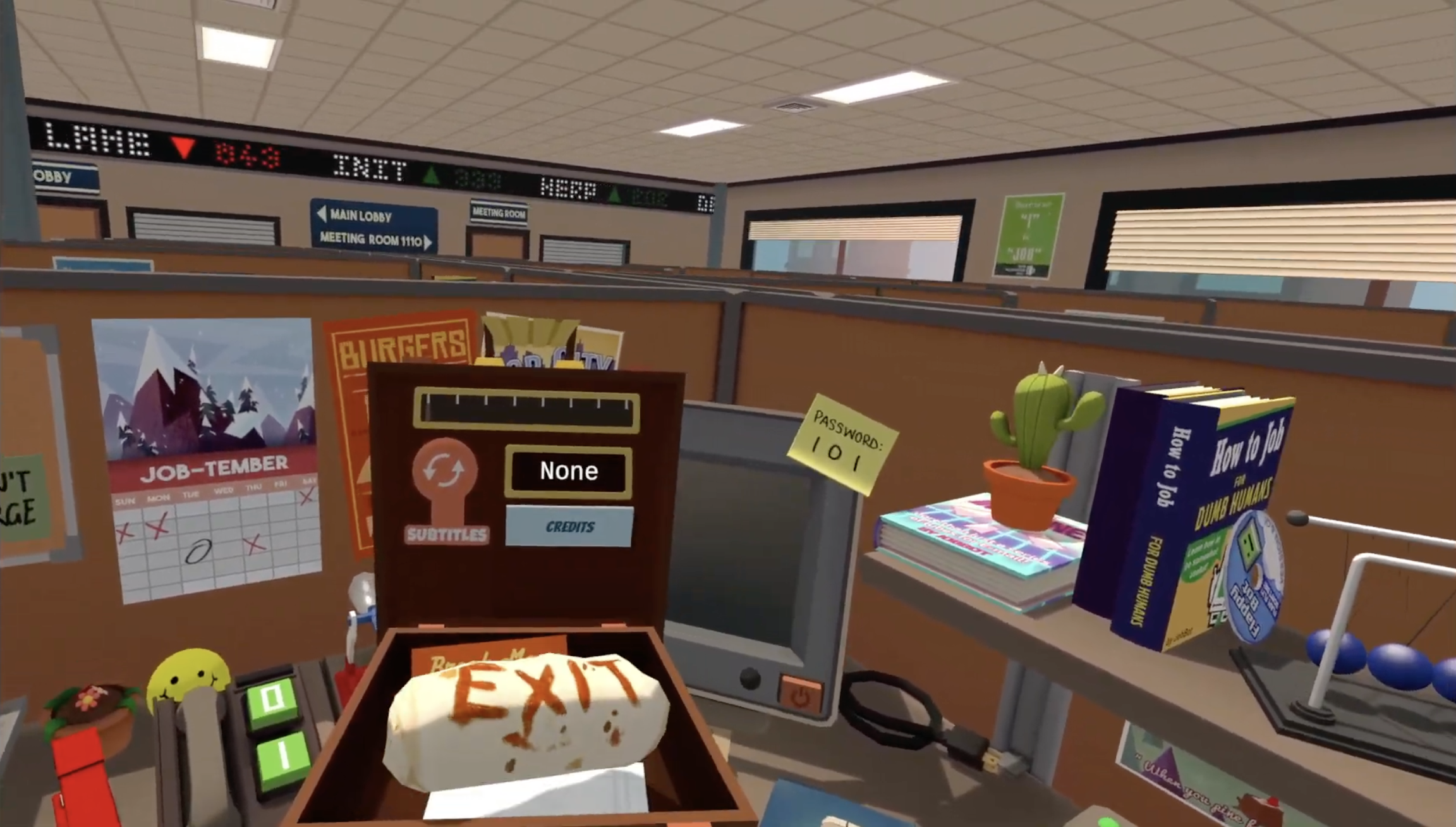}
  \caption{Briefcase with Burrito to Return to Lobby.}
  \label{fig:briefcase}
\end{figure}

Before starting the VR experience, the study runner guided the participant through two tutorials: one on how to recenter themselves in VR and another on how to navigate back to the \textit{Job Simulator} lobby from the \textit{Office Worker} and \textit{Store Clerk} jobs for the questionnaires. For the recentering tutorial, the study runner showed the participant the flat menu button they needed to click on the left-hand controller and showed them a video of which icon to select with the trigger button in order to recenter themselves (Figure \ref{fig:recentering}). The runner explained that recentering was necessary every time the participant put the headset on at the beginning of a new session to ensure consistency in both their virtual and real-world position. This also ensured that participants would not step outside the safety boundary to complete a task. For recentering, each participant stood on a tape in the center of the testing room and faced the same wall. For the lobby navigation tutorial, the study runner showed the participant the “B” button they needed to click on the right-hand controller to make the \textit{Job Simulator} briefcase appear. The runner instructed the participant to open the virtual briefcase with the grip button on their controller, pick up the burrito inside, and position it near their headset where their mouth would be to “eat” the burrito and be teleported back to the \textit{Job Simulator} lobby (Figure \ref{fig:briefcase}). The runner also showed the participant the corresponding tutorial video to act as a visual aid for the participant to understand what it looks like to navigate back to the lobby and what each button press and motion did in VR. Finally, the runner showed the participant the trigger and grip button, telling them that most of their VR interactions would be completed by squeezing the grip button to interact with virtual objects and that either the trigger or grip button could be used to respond to the virtual questionnaires. 

Once the participant verbally confirmed they understood the tutorials, the runner directed them to stand on the tape in the center of the testing room and face the wall. The runner gave instructions on how to put on and adjust the first headset. Half of the participants started with the Meta Quest 3S and the other half started with the VIVE XR Elite. The Meta Quest 3S has a knob in the back to tighten or loosen fit on the participant’s head as well as adjustable lenses that can be pinched in or out to better match the participant’s interpupillary distance (IPD) if necessary. The VIVE XR Elite similarly has a knob to tighten or loosen fit. It also has a dial at the bottom edge of the headset to fine-tune lens distance to better match the participant’s IPD if necessary. For the VIVE XR Elite specifically, we recommended that the participant position the back of the headset higher than the front of the headset to prevent the face gasket from slipping off and causing a gap revealing the real world, which could be distracting. Once the headset was positioned comfortably, the study runner handed the participant the corresponding controllers and asked them to recenter themselves while standing on the tape and facing the wall.

The runner informed the participant that they could only answer questions in the first session and not later sessions. They also encouraged the participants to respond to the questionnaires as accurately as possible.
 
Once the participant was ready, the study runner would press the “Start Recording” button in CLOVR. In the first session, the runner directed the participant to pick up the \textit{Office Worker} cartridge, insert it into the machine, and pull down the lever. They then instructed the participant to follow the instructions on the virtual board. The runner would answer questions and assist when the participant became stuck or confused. Once the participant completed all three tasks in \textit{Office Worker}, the runner would then ask them to press the B button on their right hand controller to navigate back to the \textit{Job Simulator} lobby. The runner then used CLOVR to deliver both the Fidelity-based Presence Scale (FPS) and Simulator Sickness Questionnaire (SSQ) to the user in VR. CLOVR automatically stored the participant’s responses in CSV files as well as logged the number of milliseconds between responses. This can later serve as a feature to differentiate quick responses from slower responses. 

If a participant ever selected ``severe" (level 3) in the SSQ, the study runner would immediately conclude the study for the participant's comfort. We then ejected their data from the final dataset after confirming that they intentionally selected the ``severe" response. We ran 100 individuals but ejected 5, leaving 95 participants in our dataset. 

After completing these two questionnaires, the runner would ask the participant to pick up the \textit{Store Clerk} cartridge, insert it into the machine, pull the lever, and complete the first three tasks in \textit{Store Clerk} before directing them to return to the lobby again for the second set of the same two questionnaires. At the end of each session, the study runner would ask the participant to take off the headset. They would then be offered a break or the option to put the headset back on to continue onto the next session if they so wished. Each time the participant put the headset back on, the runner would instruct them to recenter themselves again while standing on the same tape and facing the same wall as before. After three sessions in the first headset, the study runner would then switch to the second headset for the remaining two sessions. For each participant, we collected five videos, one for each session, where a session is defined as one iteration of the procedure starting when a user puts on the headset and ending when a user takes off the headset.

At the end of the study session, the participant was given a compensation form and ledger to sign before being compensated with a \$25 Amazon gift card.

\section{Dataset Attributes}

Our dataset consists of 95 participants (41 females, 53 males, 1 other). Half of the participants started with the Meta Quest 3S and ended with the VIVE XR Elite (21 females, 26 males) and the other half started with the VIVE XR Elite and ended with the Meta Quest 3S (20 females, 27 males, 1 other). There are 18 participants with partial data loss due to technical issues. The remaining 77 participants have complete session data. The age range of the dataset is from 18 to 39. The median age is 21 and the average age is 23. 90 participants were right-handed, 1 was left-handed, 2 were ``either", and 1 declined to respond. In general, the entire study took around 90 minutes. However, the total session length varied based on the participant’s prior VR experience, desire to take breaks, speed of learning, whether they engaged in off-task activities in the game, and whether the study runner had to troubleshoot technical issues. We noticed a general trend of sessions becoming shorter as participants moved through the tasks faster due to familiarity, with a slight increase from session 3 to session 4 which is when the devices switch. Table \ref{tab:session_times} displays the average length of each session. 

\begin{table}[h]
\centering
\begin{tabular}{lc}
\hline
 & \textbf{Average Time} \\
\hline
Session 1 & 13:32 \\
Session 2 & 9:07 \\
Session 3 & 7:57 \\
Session 4 & 8:16 \\
Session 5 & 7:35 \\
\hline
\end{tabular}
\caption{Average Session Times}
\label{tab:session_times}
\end{table}

Each participant's data is stored in a folder labeled with the participant's unique, random five digit identifier (e.g. 10340). This folder contains a subfolder for each session, making five subfolders in total. Within each of those session folders, there are four categories of outputs generated by CLOVR. First is a \texttt{Poses} folder contains a timestamped CSV file which records the number of active devices, the 3D position ($x, y, z$), quaternion rotation ($x, y, z, w$), linear velocity ($x, y, z$), and angular velocity ($x, y, z$) of the HMD and both controllers, as well as the number of interactions at each timestamp. Second, a \texttt{Questionnaires} folder stores two timestamped CSV files corresponding to the FPS and SSQ, each consisting of two rows reflecting the twice-per-session administration of each questionnaire. Third, a \texttt{Videos} folder contains an MP4 recording of the session. Finally, a \texttt{Project\_Properties} CSV file captures metadata about the recording session itself.

\section{Future Work \& Impact}

As we look to the future, recognizing the impact of our work in the next 25 years stems from envisioning the broader effects of having many more publicly available VR datasets with large numbers of users, multiple devices per user, multiple applications, multiple domains per user, multiple unique tasks, and multiple sessions. With more users to train on, machine learning models become more accurate in identifying and authenticating users as well as predicting attributes like gender, age, hand-dominance, and more. This can allow for more personalized experiences, better security, and seamless authentication. Identifying individuals based on motion may be transferable beyond VR -- notably, since augmented reality (AR) and AI-powered wearable technology like the Meta Ray-Ban Display will likely become prevalent within the next 25 years, individuals whom the user views in real-life via AR may also be identified \cite{rajaram_exploring_2025, rajaram_privacy_2025}. This implication emphasizes the need for advanced privacy and security protection techniques, such as obfuscation \cite{nair_going_2023, nair_deep_2024}. However, the same datasets that trigger this need can simultaneously provide a greater corpus of data for the development of such defenses. Data across multiple devices serve to enable cross-device training and testing, which can inform the design of commercial devices. As more applications and domains become represented, models will become more generalizable and ecologically valid. With more tasks, models will become more context-aware, which is especially important for the development of intelligent extended realty devices. Furthermore, datasets with multiple sessions allow for predictions of usability, user experience, performance, and learning retention over time, which can afford more intuitive user experiences, reduced simulator sickness, increased sense of presence, enhanced educational and training scenarios, and more.

Our open-source dataset serves as a significant launching point towards these future visions, providing a large set of training data with multiple devices, domains, tasks, and sessions per user. Within the next year, we intend to work on machine learning applications like personal identification, gender identification, and performance prediction, as these are some of the first steps towards creating the accurate, generalizable, and human-centric models of tomorrow.

\bibliographystyle{ACM-Reference-Format}
\bibliography{FutureHCIBib}

\end{document}